# A Quick Study of the Characterization of Radial Velocity Giant Planets in Reflected Light by Forward and Inverse Modeling

Mark Marley
Roxana Lupu
*NASA Ames Research Center*

Nikole Lewis
*Space Telescope Science Institute*

Michael Line
Caroline Morley
Jonathan Fortney
*University of California at Santa Cruz*

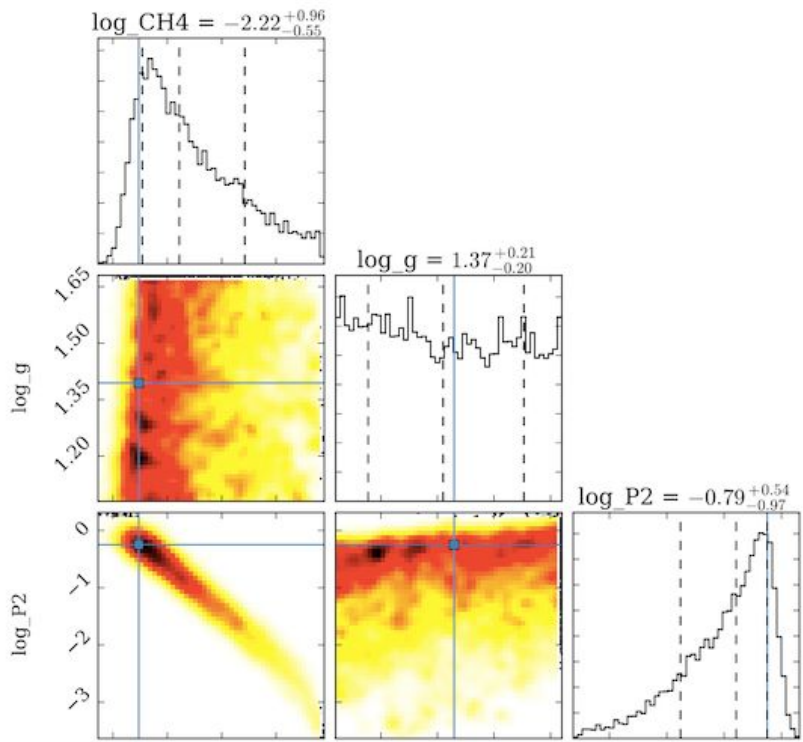

# Executive Summary

We explored two aspects of the problem of characterizing cool extrasolar giant planets in scattered optical light with a space based coronagraph. First, for a number of the known radial velocity (RV) giants we computed traditional forward models of their atmospheric structure and clouds, given various input assumptions, and computed model albedo spectra. Such models have been computed before, but mostly for generic planets. Our new models demonstrate that we can safely expect an interesting diversity of planetary spectra among those planets that are favorable for direct detection. Second, we applied a powerful Markov Chain Monte Carlo (MCMC) retrieval technique to synthetic noisy data of cool giants to better understand how well various atmospheric parameters—particularly abundances and cloud properties—could be constrained. This is the first time such techniques have been applied to this problem. The process is time consuming, so only a dozen or so cases could be completed in the limited time available. Nevertheless the results clearly show that even at S/N ~ 5, scientifically interesting and valuable conclusions can be drawn about the properties of giant planet atmospheres from noisy spectra. Further retrieval studies are clearly warranted and would be valuable to help guide instrument design decisions.

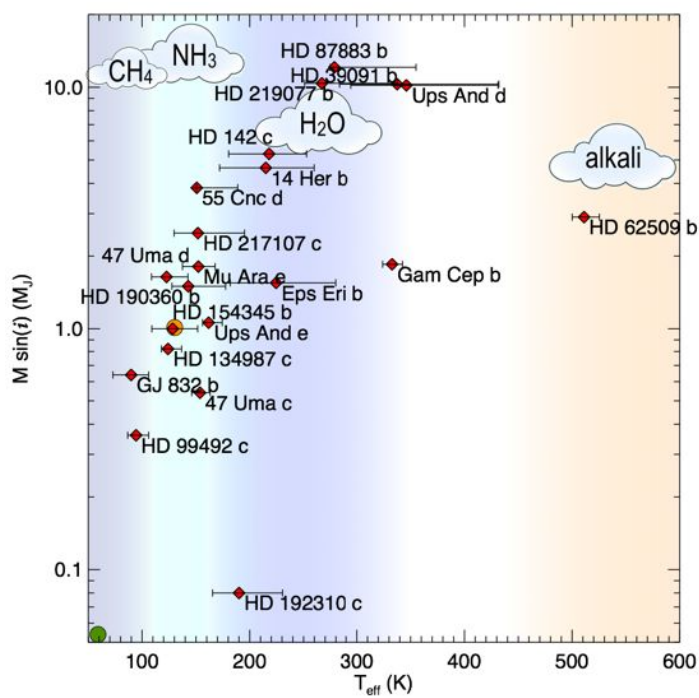

**Figure 1:** $M \sin i$ and ranges of estimated $T_{eff}$ of a selection of announced RV planets that are favorable for direct imaging. The orange circle represents Jupiter while the green one hints at Uranus which actually falls below the lower axis. Estimated $T_{eff}$ computed from planet orbits, Jupiter's albedo, and estimated internal heat flows given available estimates of the primary stars' ages. Bands show major cloud species expected in various ranges of $T_{eff}$. The existence of two of the planets shown, Ups And e and Eps Eri b, is controversial.

An example of the diversity of the known RV planets favorable for direct imaging is shown in Figure 1. In this figure the known $M \sin i$, measured by RV methods, is plotted against estimated effective temperature in order to understand the phase space of atmospheric properties that might be expected among the favorable planets. Vertical color bands show the approximate ranges over which various atmospheric compounds form clouds. The key takeaway of this figure is that while many Jupiter and Saturn-like worlds, with ammonia clouds, are expected, many planets with water, alkali, and even methane clouds may be observed.

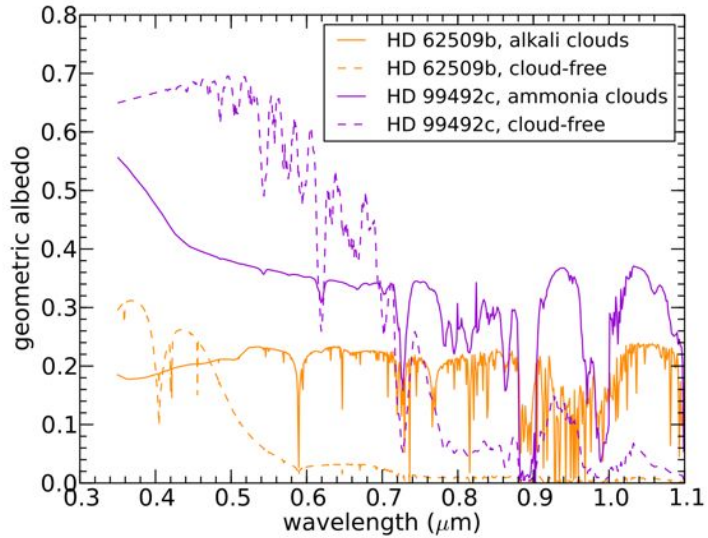

**Figure 2:** Model albedo spectra for two of the planets shown in Figure 1, HD 62509b and HD 99492c. For each planet two model cases are shown, one with and one without the clouds expected from equilibrium condensation chemistry considerations. Absorption bands seen in the HD 99492c model are methane while Na and K features are also prominent in HD 62509b.

Figure 2 shows model spectra we calculated for two of the planets shown in Figure 1. We find that HD 62509 b (Pollux b) is warm enough that the only likely atmospheric cloud condensate would be clouds of alkali species like $Na_2S$. Conversely HD 99492c is cold enough that thick ammonia, and possibly even methane, clouds are expected. The figure compares computed albedo spectra with and without the expected clouds. In the main text we present additional model spectra for various other assumptions, including gravity and atmospheric metallicity. Results such as these highlight the diversity of possible spectra. Distinctive differences, diagnostic of important atmospheric processes, between the spectra of known planets can clearly be expected.

The main thrust of our quick study, however, was not forward modeling but retrievals. We simulated observations by adding random and correlated noise to model albedo spectra spectra of solar and extrasolar giant planets and then attempted to retrieve atmospheric abundances and cloud properties. For these initial tests we constructed a highly idealized model of a scattering haze overlying an opaque cloud deck. The model is only a first attempt and unquestionably can be improved, but it was adequate for the purposes of this quick study. Our goal was not so much to verify retrieval of known quantities, but rather to determine if consistent results for scientifically interesting quantities (abundances, cloud properties) could even be obtained given the likely quality of data from a space based coronagraph studying giant planets in reflected light.

We used our sophisticated forward model albedo code and MCMC retrieval methods to explore these issues. We retrieved the model cloud properties, the atmospheric methane abundance, and the planet gravity, giving a total of 9 parameters. We focused on Jupiter, Saturn, Uranus, and HD 99492c and constructed synthetic datasets with S/N = 5, 10, and 20 and speckle noise correlation lengths of 25 and 100nm. We found that such retrieval methods could reliably infer methane abundances to within a factor of ten of the true value in most cases and could often accurately constrain cloud scattering properties, thus providing a clue to the cloud composition. Gravity, however, is not well constrained by optical spectra. For a case with spectra of S/N = 10 and gravity assumed to be known by other means to within a factor of two, the methane abundance was matched to within a factor of 3 of the known value, pointing to the value of observing planets with known masses. This removes an important source of degeneracy and allows much greater precision in the inference of atmospheric abundances. Furthermore, cases in which the cloud model was inadequate are readily apparent in the MCMC result output. In a real application we would use bayesian hypothesis testing procedures to find the optimal parameter set justified for the data.

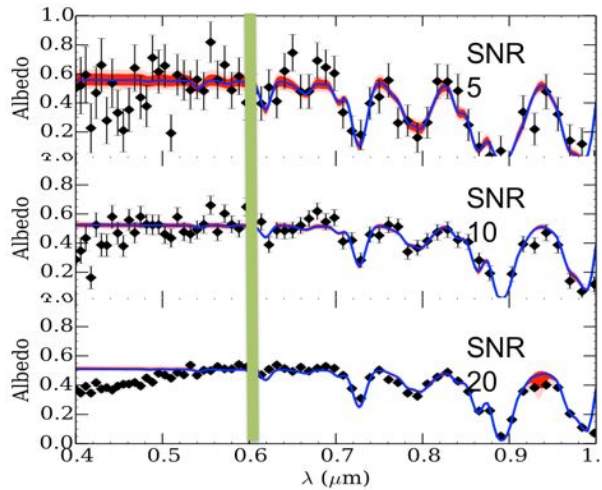

**Figure 3:** Jupiter albedo spectra (Karkoschka 1998) presented at three SNR and spectral resolution R=70. Range of the best retrieved spectra are shown by the red and blue curves. Only data longward of the vertical green line were used for the retrievals. The departure of the model from the data in the blue reveals the presence of high altitude dark hazes in Jupiter's atmosphere that are not present in the simple model used for the retrieval tests.

In summary, our quick study on the characterization of extrasolar giant planets in reflected light found or confirmed previous results:

- Great diversity is seen in the predicted spectra of the best direct imaging giant planet targets.
- Clouds, gravity, and atmospheric metallicity all play important roles in controlling planet spectra
- Retrieval methods using simple, gray cloud models can be applied to optical spectra of exoplanets to retrieve molecular abundances and cloud properties
- Unsurprisingly the best results (most accurate retrievals) are obtained at the highest SNR and the shortest speckle noise correlation length
- Noise correlation length (25 vs 100nm) is not a crucial parameter at SNR = 5, but does matter at higher SNR where the longer length appears to degrade the fidelity of the retrievals.
- Atmospheric abundances are best constrained with the planet gravity is best constrained. Thus direct imaging observations of radial velocity planets are extremely valuable as the mass can be constrained by astrometric imaging, substantially reducing the uncertainty on the gravity.

The remainder of this report presents figures supporting these conclusions and describes the methods used to reach these findings.

# Forward Model Results

Our forward model for computing giant planet atmospheric structures and albedo spectra was originally developed by Marley & McKay (1999) and Marley et al. (1999) and is based on the methods of McKay et al. (1989). The albedo code was subsequently modified and improved by Cahoy et al. (2010). The albedo model takes as input the exoplanet model's gravity and atmospheric temperature, pressure, composition, and cloud properties that are generated within the radiative–convective model. These parameters can be made variable for retrieval purposes.

Here we show a sampling of these results.

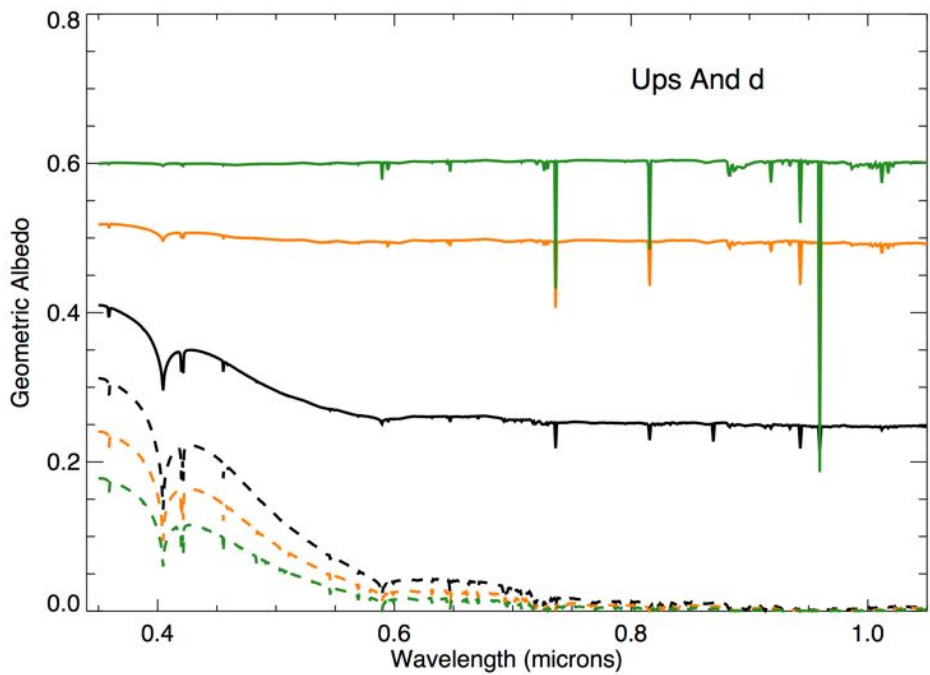

**Figure 4:** Geometric albedo spectra for the planet Ups And d (g=250 m/s, a=2.55 AU, $T_{int}$=280 K) assuming 1x (black), 3x (orange), and 10x (green) solar metallicity compositions. Sold lines represent models in which water clouds are allowed to form while dashed lines represent cloud-free models. Tenuous water clouds form in the 100 to 1 mbar region of the model atmospheres at pressures that decrease with increasing metallicity.  Increases in the assumed value of the internal heat flow, parameterized here as $T_{int}$, would increase the planet's atmospheric temperature beyond where water clouds could condense and possibly result in spectra more akin to those from the cloud free models.  Regardless, changes in assumed atmospheric metallicity produced marked difference in the resulting spectra.

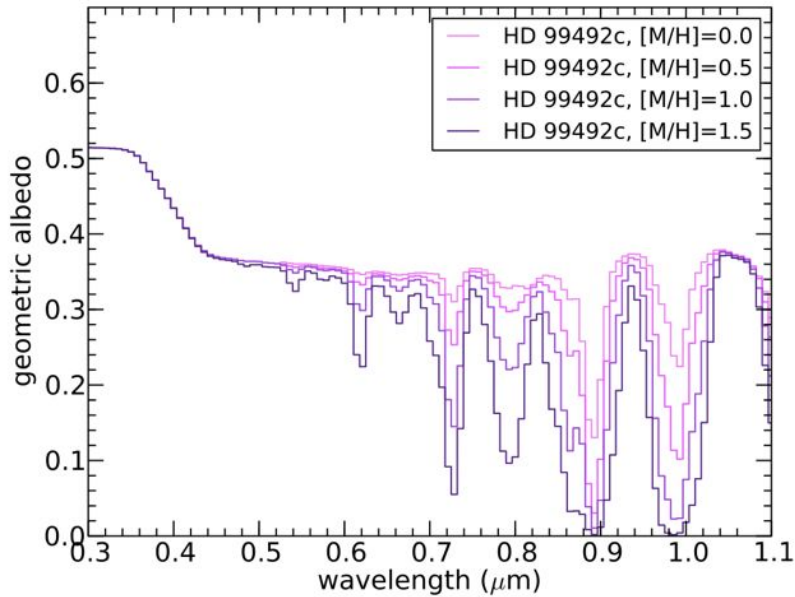

**Figure 5:** Model spectra at R=70 for HD 99492c at four metallicities, ranging from solar to 30x solar. With increasing metallicity the atmospheric methane bands deepen and weaker bands become more apparent. Thus it is crucial to detect multiple methane bands of varying strengths so that sufficient dynamic range is available to constrain a variety of atmospheric methane abundances.

## Retrieval Methods

*Overview*

For the retrieval tests we used two types of input data, solar system giants and model planets. The solar system albedo spectra are those of Karkoschka (1998). The model planet we used was HD 99492c. All of these objects have methane dominated optical reflection spectra. We did not have time to implement retrievals for warmer objects with water or alkali features and this would be an excellent future extension. For the model planet we first computed a forward 1D radiative-convective equilibrium model incorporating our sophisticated cloud model (Ackerman & Marley 2001). This model computes a self-consistent cloud with vertically varying abundances and particle sizes of each condensible species. We then input this model into our forward albedo model to produce an albedo spectrum comparable to the solar system data.

For each of the selected albedo spectra (Jupiter, Saturn, Uranus, and HD 99492c) we then modeled the instrument response to produce simulated data. The details of the noise model, which was developed by Roxana Lupu in collaboration with Wes Traub and Tom Greene, are presented in Appendix I. Note that we had originally expected to receive a noise model from the project. Since this ultimately was not forthcoming we proceeded with the development of our own model. This step took time away from other retrieval efforts. A key aspect of the noise model is the correlation length over which noise in separate spectral intervals in the IFS is correlated. We considered two cases, 25 and 100 nm. For each planet we considered 3 signal-to-noise ratios, 5, 10, and 20, for a total of six cases.

Note that for our purposes here we modeled the planet as observed at full phase, utilizing the observed and computed geometric albedo spectra. In the future we will consider the more realistic case of observations at partial phase. However we did not introduce that complexity for this initial quick study. For the atmospheric clouds and hazes we constructed a very simple 2 layer cloud model defined by 7 model parameters, including the single scattering albedos of the cloud and haze. All parameters were treated as grey over the wavelength range. Details of the cloud model are described below.

In addition to the cloud properties we retrieved the gravity and atmospheric methane mixing ratio. We allowed gravity to vary because in the realistic case neither the size of the planet nor the planetary mass will be known precisely. We define the atmospheric methane mixing ratio, $f_{CH4}$, as the volume mixing ratio of methane. Since in a giant planet atmosphere 98% of the atmospheric constituents are $H_2$ and He, this uniquely defines the atmospheric methane content. Such an approach would not be possible for a terrestrial planet of course. If the planet were composed predominantly of heavier gasses then, by the mass-radius relationship, the planet would be substantially smaller and darker, again ruling such cases out.

We allowed an exceptionally large range of gravities to be tested by the retrievals. In a realistic case the planet mass will be known to substantially better than a factor two by the orbital astrometry solution. From the mass radius relationship for gas giant planets and albedo scaling arguments (below) the radius will be known to within 50%, which dominates the gravity uncertainty through $R^2$. Thus for a Jupiter twin the gravity ($g = 25$ m s$^{-2}$) would plausibly be known to be $< 100$ m s$^{-2}$, not $< 1000$ m s$^{-2}$, which is the constraint placed in most of the retrievals shown here. This turned out to be very important as, all else being equal, a large methane mixing ratio is required at high gravity to produce the same depth absorption feature as a lower abundance at lower gravity. The high gravity solutions were responsible for pulling the best fits to higher mixing ratios than the known values for solar system giants.

***Methods***
Based on our experience and the results of Cahoy et al. (2010), we expect that the most relevant model parameters will be methane abundance, surface gravity, and cloud properties. The simplified model used in the present study has a constant methane abundance throughout the atmosphere. The only other components are $H_2$ and He in primordial solar ratio. The mean molecular weight of the atmosphere is calculated using these 3 components. The code is set up such that additional opacity sources can be added in the future. We characterize the cloud structure with 2 cloud layers and 7 parameters (Figure 6), which is a slight modification of those used in the classic analysis of Jupiter's atmosphere by Sato & Hansen (1979 ).

The specific gray cloud parameters include the pressure at the top of the lower cloud and its single-scattering albedo ($\varpi$), the pressure difference between the top of the lower cloud and the bottom of the upper cloud (dP1), the pressure difference between the top and the bottom of the upper cloud (dP2), the single-scattering albedo ($\varpi$), asymmetry factor ($\breve{g}$) and total optical depth ($\tau$) of the upper cloud. Hazes that lower the albedo in the short-wavelength part of the spectrum are not taken into account. These effects become more important below 0.5 µm, and are unlikely to affect the region of interest for this study (0.6-1 µm).

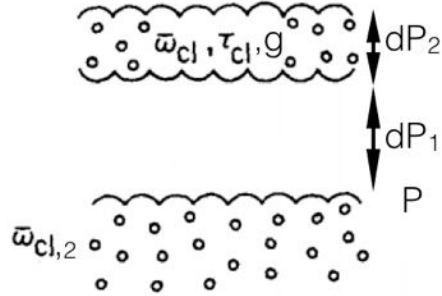

**Figure 6.** Graphical representation of our 2-cloud model.

This model framework thus includes a **total of nine forward model parameters:**
   $f_{CH4}$, $g$
   $dP_1$, $dP_2$, $\tau$, $\varpi_1$, $\breve{g}$ for the upper cloud
   $P$, $\varpi_2$ for the bottom cloud

To simulate a spherical planet, we cover the illuminated surface of a sphere with many plane–parallel facets, where each facet has different incident and observed angles, as shown in Figure 7. In our simplified approach we only consider the 0-degree phase angle (face-on), in which case the observer and the source are collinear ($\mu_0=\mu_1$). Following the approach of Horak (1950) and Horak & Little (1965), we use two-dimensional planetary coordinates and Chebyshev–Gauss integration to integrate over the emergent intensities and calculate the albedo spectra. The radiative transfer is performed line-by-line for each of the points sampling the planetary disk. Our most accurate albedo code uses 1000 angle combinations and different incident and observer angles, but we have reduced this number to 100 for the current study. In this case, the albedo code takes about 4s to run, which is reasonable to use in combination with a Monte Carlo Markov Chain.

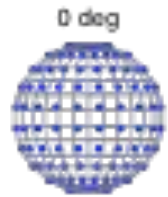

**Figure 7.** Each dot represents one plane–parallel albedo spectra model. A two-dimensional Chebyshev–Gauss integration over all dots is performed to calculate the albedo spectra. For simplicity, the current study was restricted to 0 degree phase angle.

We use the affine invariant ensemble MCMC sampler, EMCEE (Goodman & Weare 2010, Foreman-Mackey et al. 2013) to compute the optimal set of parameters and their uncertainties. This approach is favorable over others as it permits efficient sampling of highly correlated, non gaussian, and high-dimensional parameter spaces. It is also very readily scaleable to multi-processor computing.

*Results*

We have applied our albedo retrieval method to a set of 24 cases, comprising 6 combinations of SNR (5, 10, 20) and noise correlation lengths (25 and 100 nm) for 4 model planets: Jupiter, Saturn, Uranus, and the extrasolar gas giant HD99492c. The Solar System-like planets are assumed to be at 25 pc from the Earth in each case. The HD99492c system is at 18 pc from the Sun. For brevity, we only present a representative selection of the parameter distributions obtained in this study. For each case we run a MCMC ensemble sampler with 24 walkers per parameters, for a total of 4000 steps. We select the last 1500 steps for determining the posterior probability distribution. This set contains 324000 samples and is statistically independent from the starting conditions.

**Jupiter**

Figures 8-10 show the covariances of the forward model parameters for SNR of 5, 10, and 20, respectively, and a noise correlation length of 25 nm. Figure 11 shows the model fits to the spectra corresponding to all six cases. As shown in Figure 12, the widths of the marginalized probability distributions of all parameters are similar, regardless of SNR. However, the peak of the probably distribution can change, depending on the noise on the simulated data (see the black and blue lines in the first panel). The methane abundance is correlated with the surface gravity, but is still constrained to within factors of 20-30 over 2 orders of magnitude in *g*. If the value for g is constrained to within a factor of 2, the peak of the posterior probability distribution for the $CH_4$ abundance (red curve in the bottom row, first panel) matches the measured $CH_4$ abundance in Jupiter (Sato & Hansen 1979). Also the derived single scattering albedo of the lower cloud, 0.99, precisely matches the observed value of 0.997 (e.g., Sato et al. 2013). Figure 13 shows a cartoon of the vertical cloud structure of a Jupiter-like planet, as inferred from our simple albedo model.

We conclude that the two-layer cloud model works well for Jupiter, constraining the methane abundance to within factors of 20-30, and the single scattering albedo of the lower cloud is constrained within 0.5%. This gives us an indication for the composition of the lower cloud, since particles with high reflectivity are necessary to explain the large value of $\varpi$. We can also can place limits of the extent and position of the upper cloud, as well as its optical depth. The gravity, scattering asymmetry factor, and single scattering albedo of the upper cloud are essentially unconstrained by these data. However, for known RV planets, the mass and radius, and therefore *g*, can be constrained independently. This will considerably tighten the posterior probability distribution for the methane abundance.

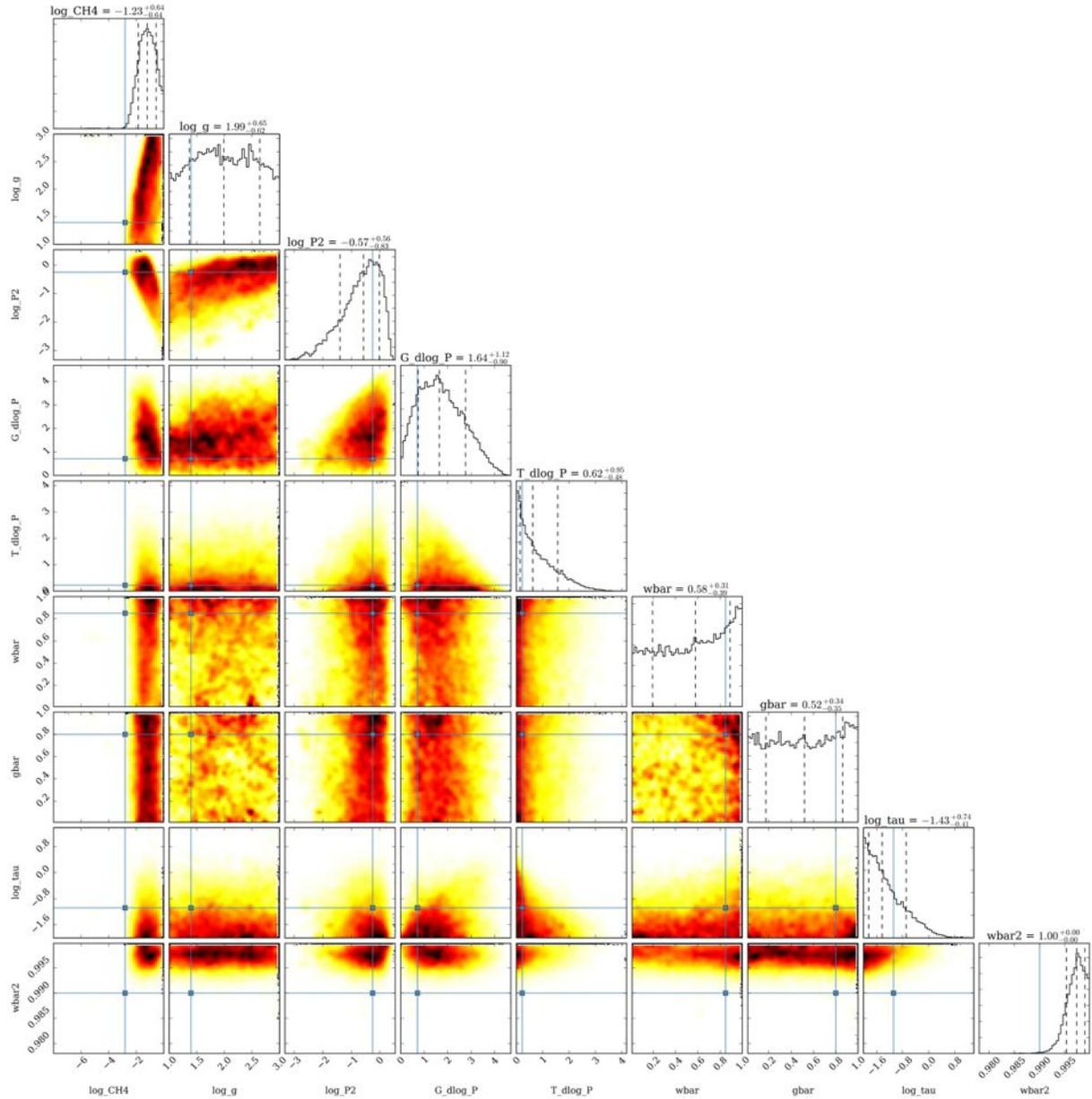

**Figure 8:** Posterior probability distributions for the albedo model parameters, as sampled by the MCMC algorithm, for the Jupiter albedo with simulated data for SNR=5 and noise correlation length 25 nm. The distributions contain 324000 samples, after excluding the burn-in phase for the MCMC chain. The darker regions represent higher probability. The marginalized probability distributions for each parameter are shown on the diagonal. The error bars and the expectation value are calculated to indicate the 16%, 84% and 50% quantiles, respectively, such that the entire confidence interval contains 68% of the values.

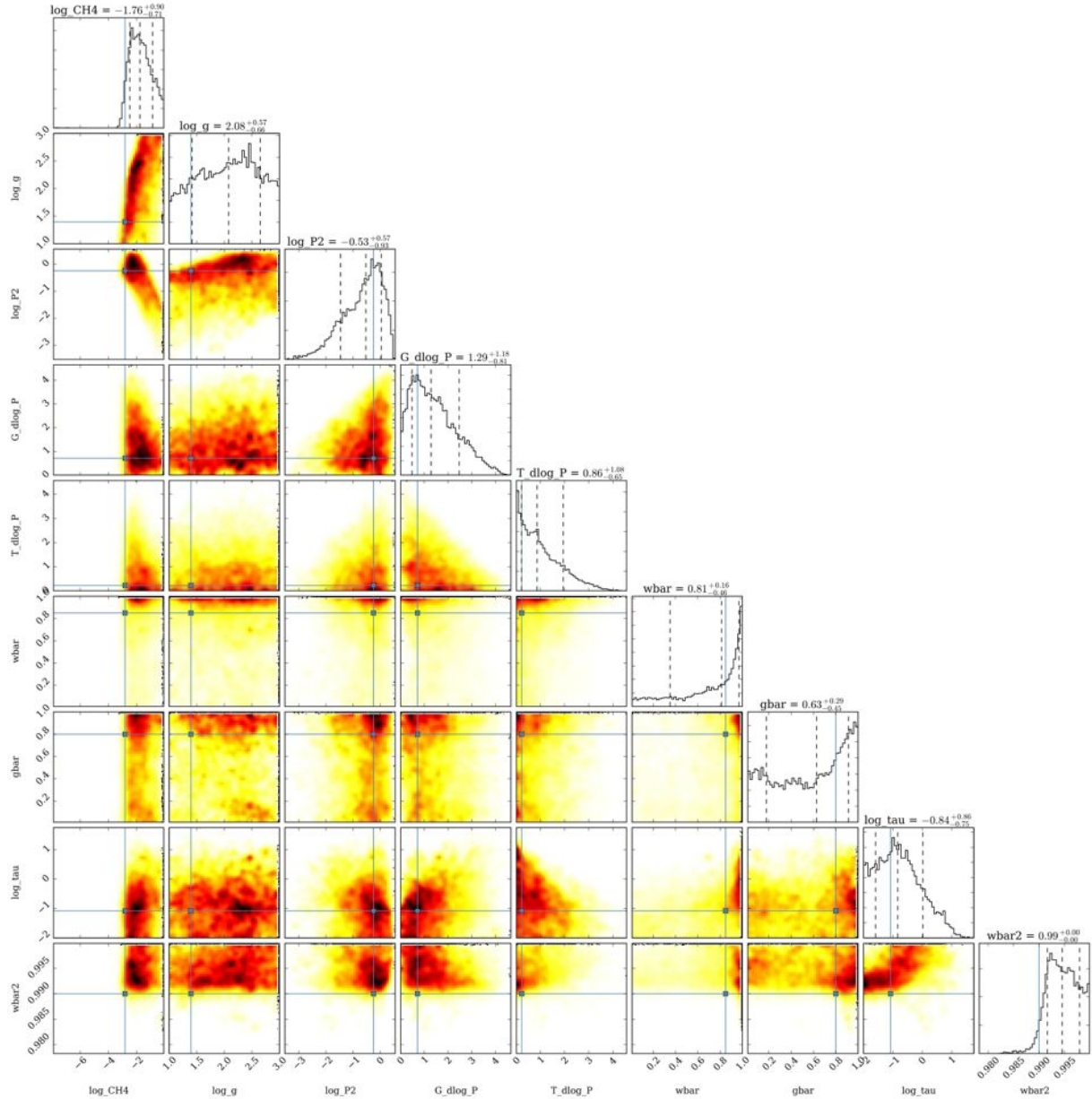

**Figure 9:** Posterior probability distributions for the albedo model parameters, as sampled by the MCMC algorithm, for the Jupiter albedo with simulated data for SNR=10 and noise correlation length 25 nm. The distributions contain 324000 samples, after excluding the burn-in phase for the MCMC chain. The darker regions represent higher probability. The marginalized probability distributions for each parameter are shown on the diagonal. The error bars and the expectation value are calculated to indicate the 16%, 84% and 50% quantiles, respectively, such that the entire confidence interval contains 68% of the values.

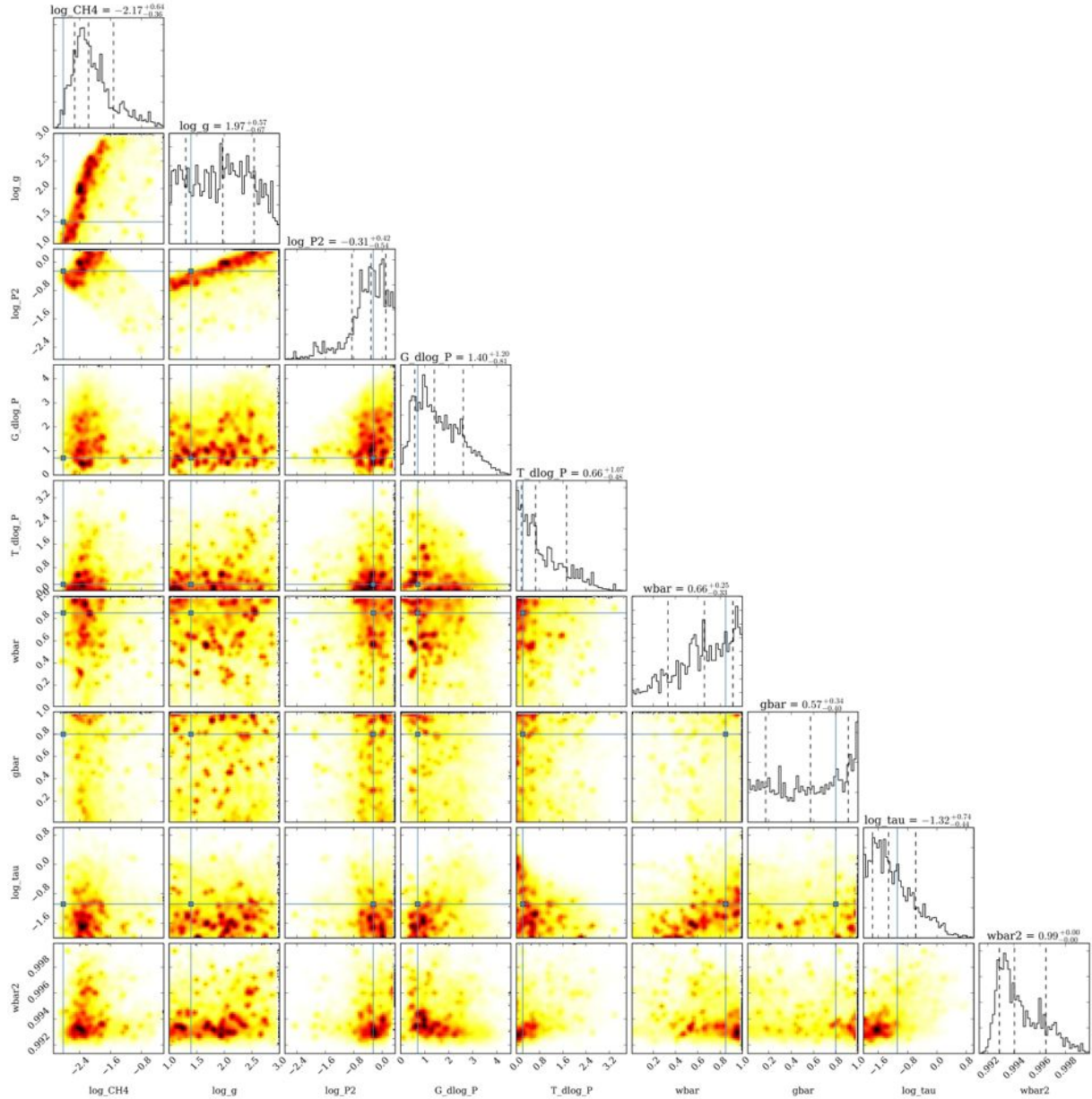

**Figure 10:** Posterior probability distributions for the albedo model parameters, as sampled by the MCMC algorithm, for the Jupiter albedo with simulated data for SNR=20 and noise correlation length 25 nm. The distributions contain 324000 samples, after excluding the burn-in phase for the MCMC chain. The darker regions represent higher probability. The marginalized probability distributions for each parameter are shown on the diagonal. The error bars and the expectation value are calculated to indicate the 16%, 84% and 50% quantiles, respectively, such that the entire confidence interval contains 68% of the values.

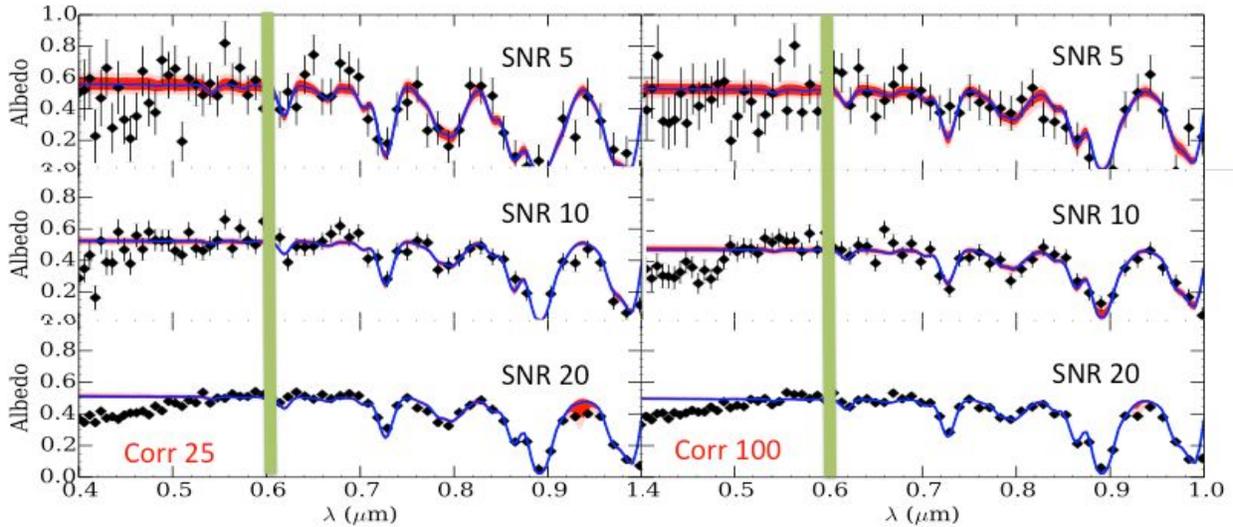

**Figure 11.** The fits to the data obtained for Jupiter by running the MCMC ensemble sampler. The model spectra are calculated for 1000 samples of the 324000 final sample chain. The blue line represents the median spectrum, the dark red the region contains the 16 to 84% percentile range of the spectra, while the light red contains the 4.5 to 95.5 % percentile range. The green lines mark the wavelength longward of which we fit the models to the data.

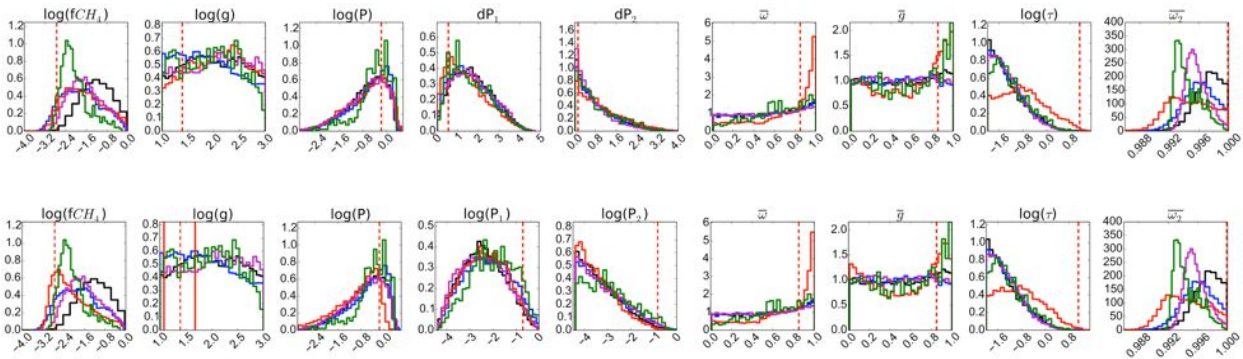

**Figure 12.** Overlaid marginal probability distributions for the albedo model parameters for the reflected spectrum of Jupiter, each color corresponding to one (SNR, correlation length) combination. The red lines indicate the position of the real values for Jupiter. The lower panel shows the change in the probability distribution for log(fCH4) after narrowing the allowed range for *g* to include the actual value for Jupiter (24.79 m/s$^2$), plus minus a factor of 2 (the red distribution). The peak of the probability distribution now matched the actual measured value for fCH4 in Jupiter: 1.8x10$^{-3}$. In the lower panel, the pressure differences (dP1 and dP2) are also converted in actual pressures at the bottom (P1) and top (P2) of the upper cloud. This can be used to get a graphical representation of the cloud structure, as shown in Figure 13.

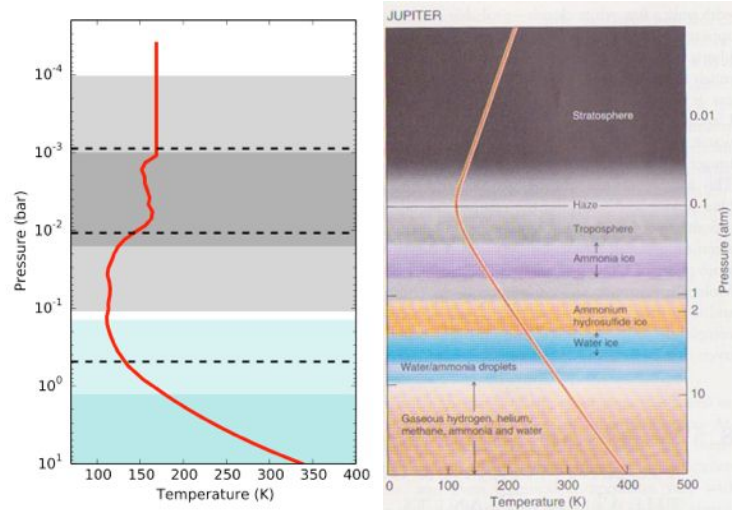

**Figure 13.** Vertical cloud structure for Jupiter, as inferred from a 7-parameter forward albedo model. The grey cover the 1-sigma confidence intervals for the upper cloud, while the light blue shows the 1-sigma confidence interval for the top of the lower cloud. The measured pressure-temperature profile for Jupiter is shown in red. For comparison, the right panel shows a textbook model for Jupiter's atmosphere.

**Saturn**

Figures 14-16 show our results for Saturn. Figure 14 is an example for the parameter covariances and marginalized posterior probability distributions. The distributions for the pressure at the top of the bottom cloud, its single scattering albedo, as well as the optical depth of the upper and lower cloud are bi-modal. The surface gravity and the scattering asymmetry factor are again unconstrained. This model only places a lower limit on the methane abundance, close to the measured value. The parameter space for the cloud parameters is roughly separated in 2 regions. In the first case the lower cloud is placed high (above 1 bar), with single scattering albedo peaked at 1, while the upper cloud is optically thin, and its single scattering albedo is unconstrained. In the second case, the lower cloud is placed roughly below 10 bars, but has an unconstrained single scattering albedo, while the top cloud is optically thick, and it's single scattering albedo is high. Essentially, the result suggests that only one cloud is "visible" in either case: the lower one in the first case, and the upper one in the second. Both cases are consistent with highly reflective layers present in the atmosphere. Only upper limits can be derived for the position and extent of the upper cloud.

Figure 15 shows the spectral fits for a noise correlation length of 25, while Figure 16 summarizes all marginalized probability distributions obtained in this study. The confidence intervals for the model parameters are similar regardless of SNR. The lower panel shows that limiting the priors on some parameters does not affect substantially the overall posterior probability distributions (red curve).

We conclude that the poor parameter constraints suggest that our model is not a good description of the data. Indeed, a three-cloud layer model would be a better description for Saturn. The power of the MCMC method, however, is that the shortcoming of the cloud model is immediately apparent and in a real use situation would motivate increasing the complexity of the model.

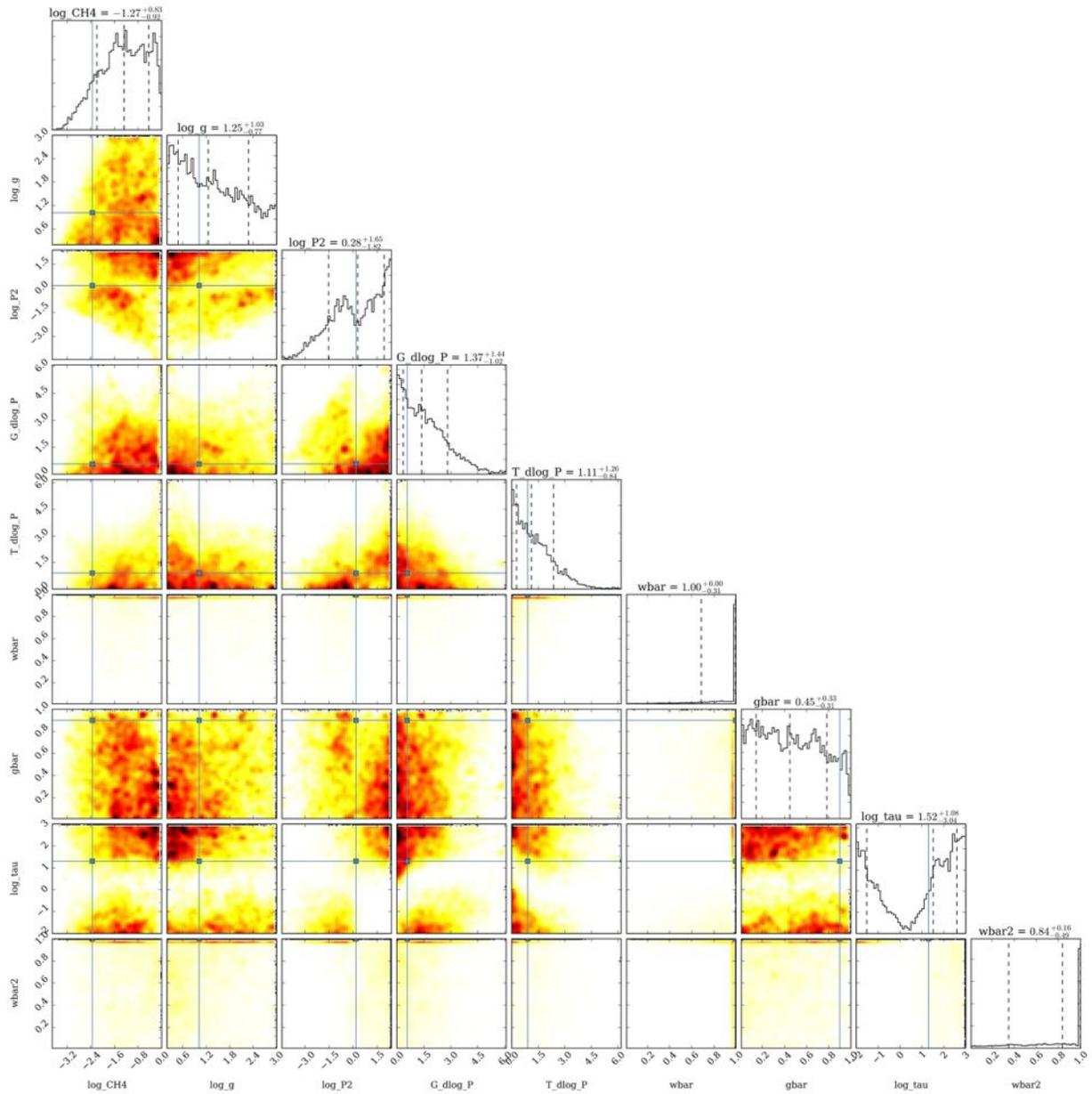

**Figure 14** Posterior probability distributions for the albedo model parameters, as sampled by the MCMC algorithm, for the Saturn albedo with simulated data for SNR=5 and noise correlation length 25 nm. The distributions contain 324000 samples, after excluding the burn-in phase for the MCMC chain. The darker regions represent higher probability. The marginalized probability distributions for each parameter are shown on the diagonal. The error bars and the expectation value are calculated to indicate the 16%, 84% and 50% quantiles, respectively, such that the entire confidence interval contains 68% of the values.

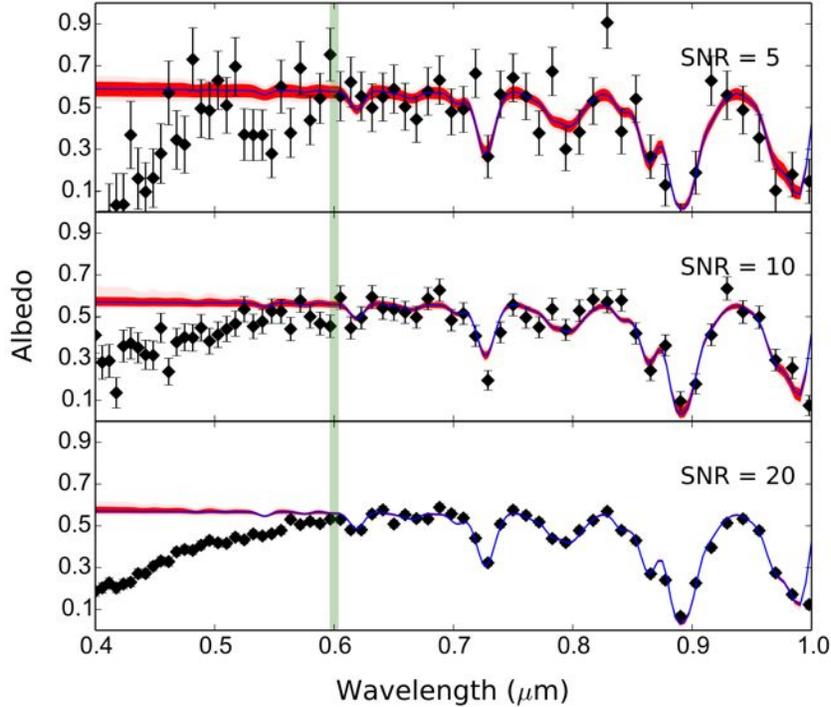

**Figure 15.** The fits to the data obtained for Saturn by running the MCMC ensemble sampler. The model spectra are calculated for 1000 samples of the 324000 final sample chain. The blue line represents the median spectrum, the dark red the region contains the 16 to 84% percentile range of the spectra, while the light red contains the 4.5 to 95.5 % percentile range. The green lines mark the wavelength longward of which we fit the models to the data. The shown models are for a noise correlation length of 25 nm and SNR of 5, 10, and 20, from top to bottom.

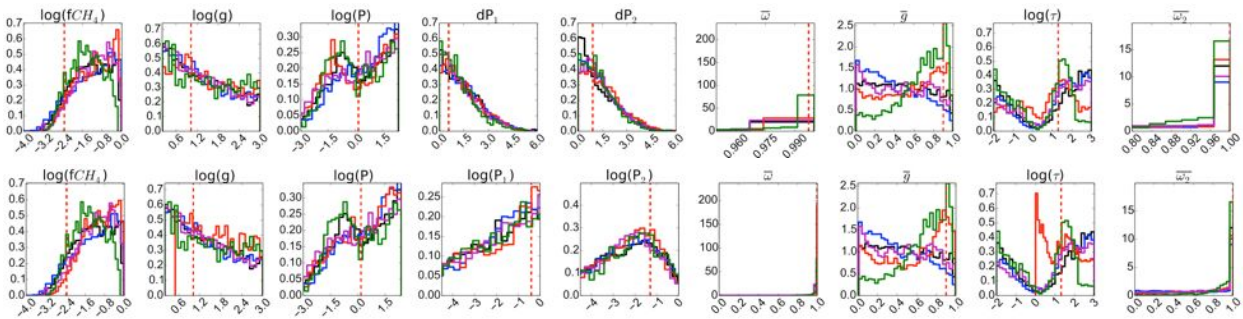

**Figure 16.** Overlaid marginal probability distributions for the albedo model parameters for the reflected spectrum of Saturn, each color corresponding to one (SNR, correlation length) combination. The red lines indicate the position of the real values for Saturn. In the lower panel, the pressure differences (dP1 and dP2) are converted in actual pressures at the bottom (P1) and top (P2) of the upper cloud. The optical depth of the upper cloud and the surface gravity have also been slightly restricted in one of the models (red line), but this does not affect the final posterior distribution of the other parameters.

**Uranus**
Figures 17-19 show our results for Uranus. Figure 17 is an example for the parameter covariances and marginalized posterior probability distributions, while Figure 19 shows the fits to the spectra as a function of SNR. The blue line indicates the methane mixing ratio below 1.3 bar, which is outside the maximum probability region. However, above this region, the methane mixing ratio is largely 0.1%, similar to the peak of our probability distribution. We obtain an upper limit for the optical depth of the upper cloud, but the other upper cloud parameters are essentially unconstrained. We also obtain a lower limit for the single scattering albedo of the lower cloud and an upper limit for its top pressure. These upper limits are consistent with the current models that place 4 cloud layers below a pressure level of ~ 1bar (Lindal et al. 1987). The marginalized probability distributions are summarized in Figure 19.

We conclude that the extremely low value for the surface gravity and the upper limits on cloud parameters suggests that the simple two-cloud model is not a good description of the atmospheric structure. Multiple ice cloud layers are known to be found in the lower atmosphere of Uranus. The discrepancies between the fit spectra and the simulated data at short wavelengths also suggest that extending the spectral range of the observations to shorter wavelengths would help characterize this type of planet. As the SNR increases, more discrepancies between the model and the data become apparent at longer wavelengths. This is an additional indication that the assumed cloud structure is inadequate. However, the derived abundance of methane is within the accepted range, and it can be further constrained with a more realistic cloud structure. As it was shown in the case of Saturn, the MCMC retrieval method is able to inform the atmospheric model selection and identify situations in which the assumed model framework is inadequate.

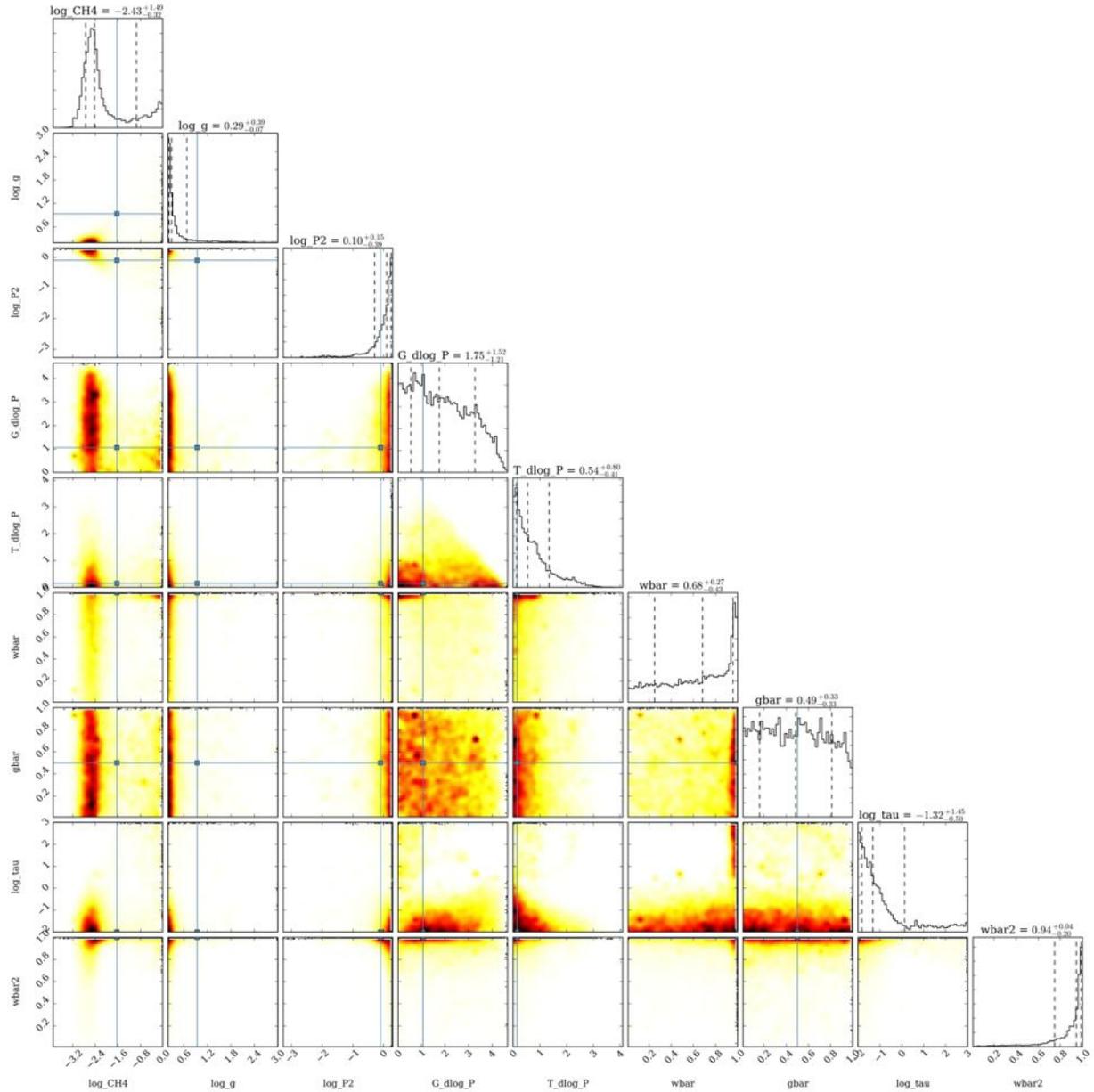

**Figure 17** Posterior probability distributions for the albedo model parameters, as sampled by the MCMC algorithm, for the Uranus albedo with simulated data for SNR=5 and noise correlation length 25 nm. The distributions contain 324000 samples, after excluding the burn-in phase for the MCMC chain. The darker regions represent higher probability. The marginalized probability distributions for each parameter are shown on the diagonal. The error bars and the expectation value are calculated to indicate the 16%, 84% and 50% quantiles, respectively, such that the entire confidence interval contains 68% of the values.

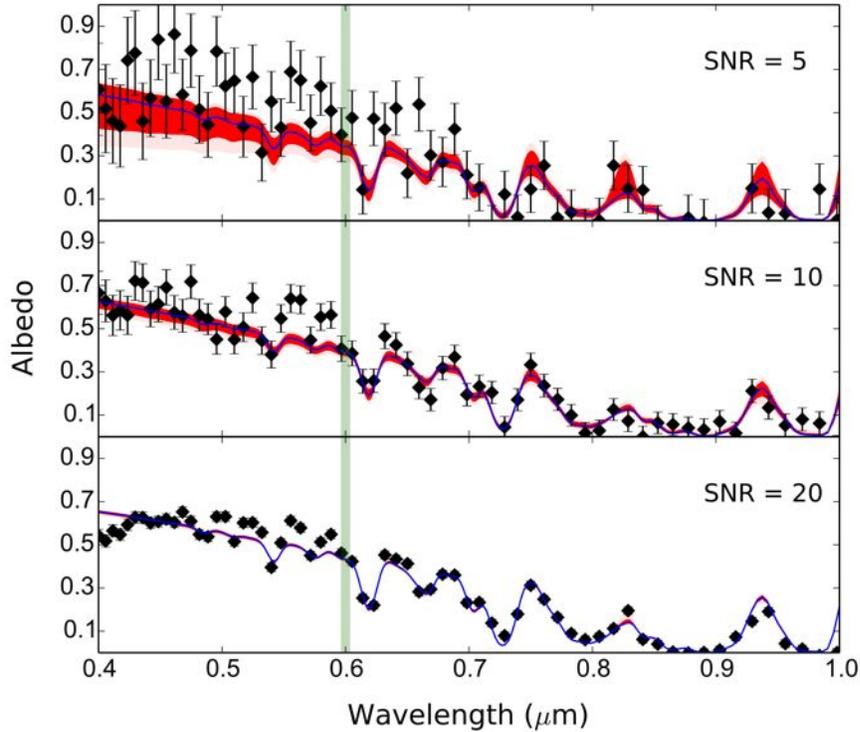

**Figure 18.** The fits to the data obtained for Uranus by running the MCMC ensemble sampler. The model spectra are calculated for 1000 samples of the 324000 final sample chain. The blue line represents the median spectrum, the dark red the region contains the 16 to 84% percentile range of the spectra, while the light red contains the 4.5 to 95.5 % percentile range. The green lines mark the wavelength longward of which we fit the models to the data. The shown models are for a noise correlation length of 25 nm and SNR of 5, 10, and 20, from top to bottom.

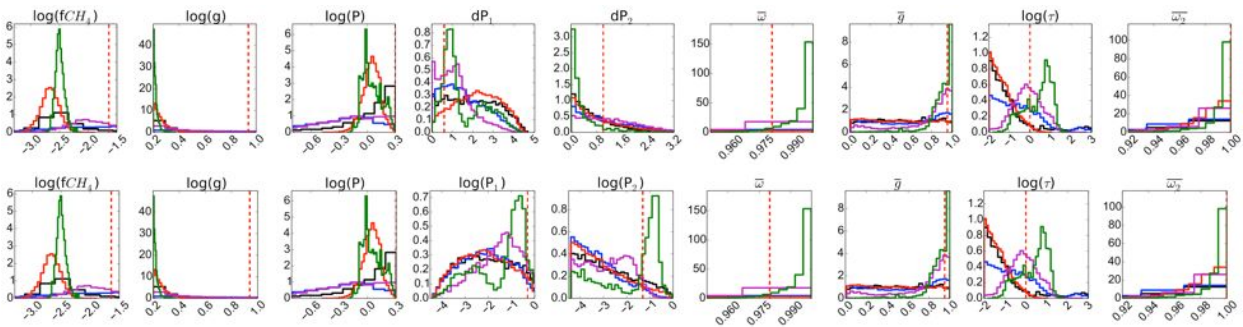

**Figure 19.** Overlaid marginal probability distributions for the albedo model parameters for the reflected spectrum of Uranus, each color corresponding to one (SNR, correlation length) combination. The red lines indicate the position of the real values for Uranus. In the lower panel, the pressure differences (dP1 and dP2) are converted in actual pressures at the bottom (P1) and top (P2) of the upper cloud.

**HD99492c**

Figures 20-22 show our results for the gas giant HD99492c. Figure 20 is an example for the parameter covariances and marginalized posterior probability distributions, while Figure 21 shows the fits to the spectra as a function of SNR. We obtain very tight constraints on the single scattering albedo of the top cloud (4%). The scattering asymmetry factor for the top cloud is correlated with its single scattering albedo. The single scattering albedo of the lower cloud is unconstrained when the optical depth of the upper cloud is large. This is to be expected, since in this case we cannot "see" the lower cloud. For low optical depths of the upper cloud, the single scattering albedo of the lower cloud would be constrained to high values, which would favor lower values for the mixing ratio of methane. We obtain only an upper bound for the position of the top of the lower cloud. The correlation between the surface gravity and the mixing ratio of methane is still noticeable, as in the case of Jupiter. The surface gravity and the asymmetry factor for the upper cloud are still poorly constrained. Figure 22 shows that the constraints for some of the parameters do increase with the SNR of the simulated data (red=SNR of 10, green = SNR of 20).

We conclude that the tight constraint on the single scattering albedo of the upper cloud and the region of the parameter space consistent with the lower cloud being obscured by the upper cloud suggests that a single cloud model would be appropriate for this type of planet. The number of parameters could be significantly reduced in this case, and therefore we could expect to better constrain the remaining parameters. Knowing the surface gravity will also tighten the constraints on the methane mixing ratio.

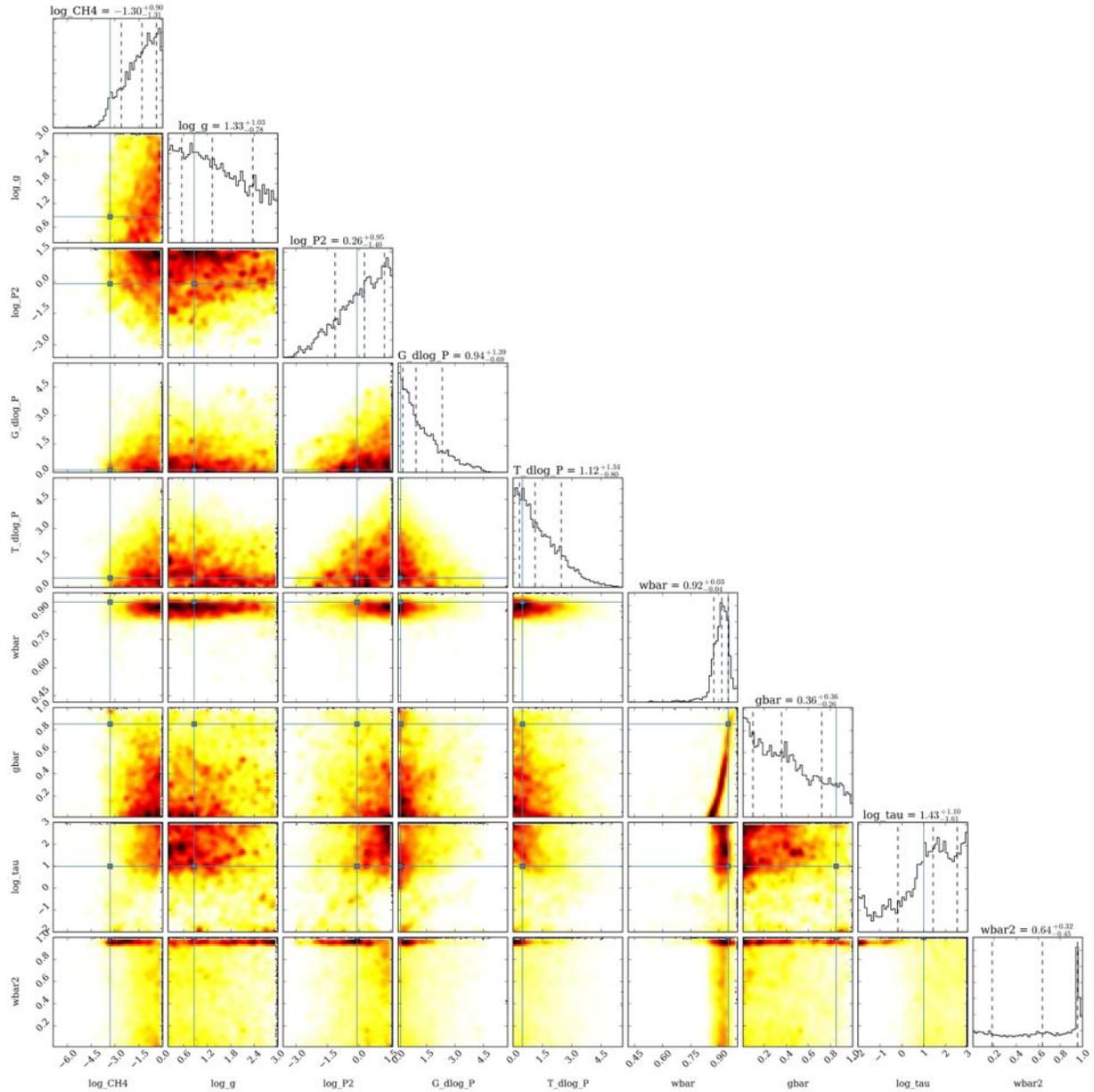

**Figure 20** Posterior probability distributions for the albedo model parameters, as sampled by the MCMC algorithm, for the Jupiter albedo with simulated data for SNR=20 and noise correlation length 25 nm. The distributions contain 324000 samples, after excluding the burn-in phase for the MCMC chain. The darker regions represent higher probability. The marginalized probability distributions for each parameter are shown on the diagonal. The error bars and the expectation value are calculated to indicate the 16%, 84% and 50% quantiles, respectively, such that the entire confidence interval contains 68% of the values.

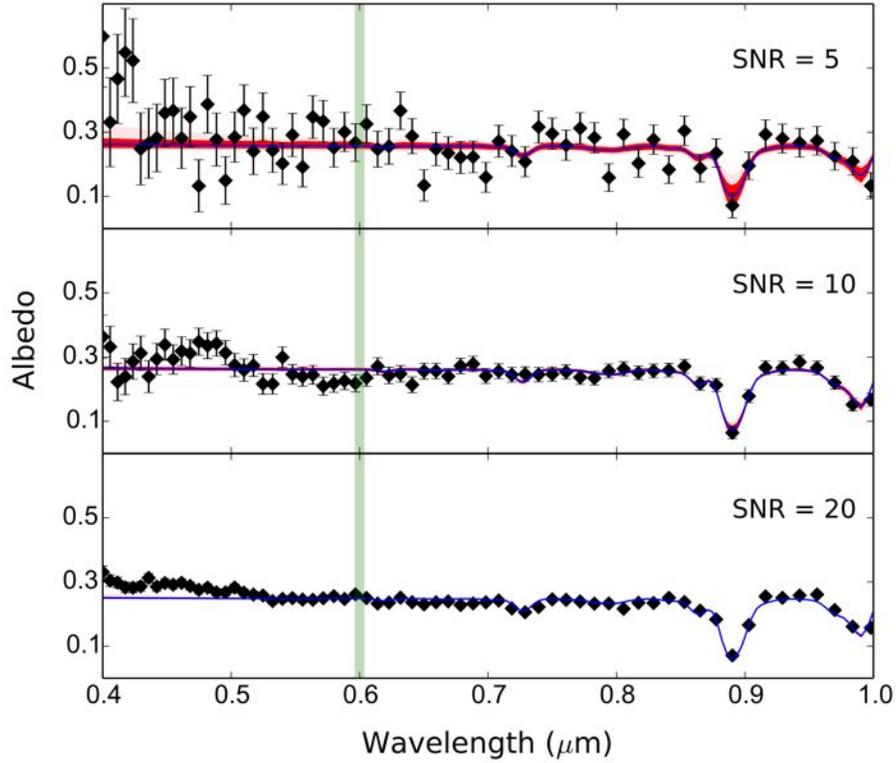

**Figure 21.** The fits to the data obtained for the gas giant HD99492c by running the MCMC ensemble sampler. The model spectra are calculated for 1000 samples of the 324000 final sample chain. The blue line represents the median spectrum, the dark red the region contains the 16 to 84% percentile range of the spectra, while the light red contains the 4.5 to 95.5 % percentile range. The green lines mark the wavelength longward of which we fit the models to the data. The shown models are for a noise correlation length of 25 and SNR of 5, 10, and 20, from top to bottom.

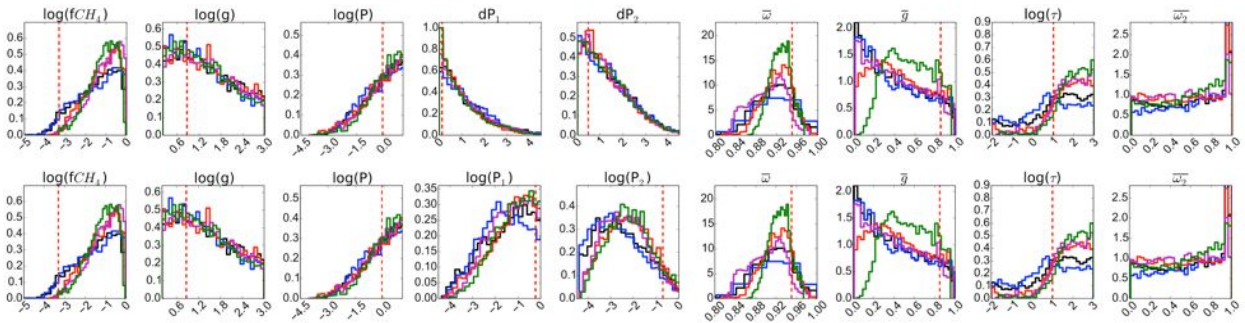

**Figure 22.** Overlaid marginal probability distributions for the albedo model parameters for the reflected spectrum of the gas giant HD99492c, each color corresponding to one (SNR, correlation length) combination. The red lines indicate the position of the real values for HD99492c. In the lower panel, the pressure differences (dP1 and dP2) are converted in actual pressures at the bottom (P1) and top (P2) of the upper cloud.

**Importance of SNR and Noise Correlation Length**

Figures 23a and b summarize the results of the MCMC modeling discussed above by showing the retrieved methane abundance and lower cloud single scattering albedo for each of the four planets at 3 SNR and 2 noise correlation length scales. The most straightforward abundance results are obtained for Jupiter, which is not surprising since the forward model was constructed with Jupiter's atmosphere in mind. For this planet the smallest range of best fitting methane abundances are found for the highest SNR model. While the influence of the spectral correlation length is less obvious, at both Jupiter and Uranus it is clear that better results are obtained with the lower correlation length. For the other planets the case is less clear. For the cloud single scattering albedo it seems that either the model framework is either appropriate and a clear result obtained, as for Jupiter and Uranus, or the model is not sufficient, as for Saturn and HD 99492c. The successful constraint of the lower cloud for Jupiter and Uranus would allow inferences on the cloud composition to be drawn. Since the cloud scattering properties are inferred from the average height of the continuum flux, the noise correlation length of the detector is not an important parameter for these model inferences.

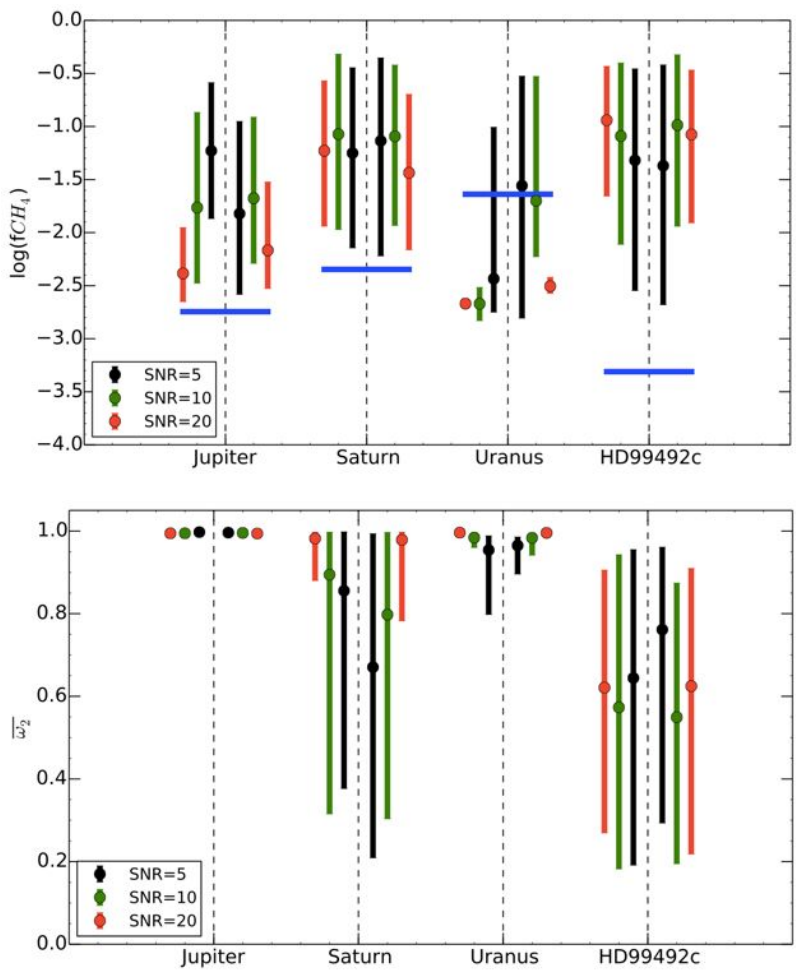

**Figure 23** (a, upper and b, lower): Range of derived atmospheric methane mixing ratio and lower cloud single scattering albedo for the four planets considered here. For each planet six model results are shown. These are for synthetic data with SNR of 5, 10, and 20 and noise correlation lengths of 25 and 100 nm. The 25 nm lengthscale are for the models to the left of the vertical dashed line for each planet while the 100 nm case is to the right. In the case of methane the blue line denotes the measured (solar system planets) or modle (HD99492c) value.

**Value of Gravity Constraint**

As discussed above, for some planets the value of the gravity will be known much more precisely than we assumed here. To better understand the value of such a gravity constraint we conducted one retrieval run for Jupiter where instead of only slightly limiting *g* to be in the brown dwarf regime or less, we assumed that *g* was known within a factor of two of the true value. The resulting posterior probability distribution is shown in Figure 24 and the methane improvement is highlighted in Figure 25. Because high gravity, high abundance cases that previously produced good data fits are now excluded, the derived abundance now more closely reproduces the known value for Jupiter.

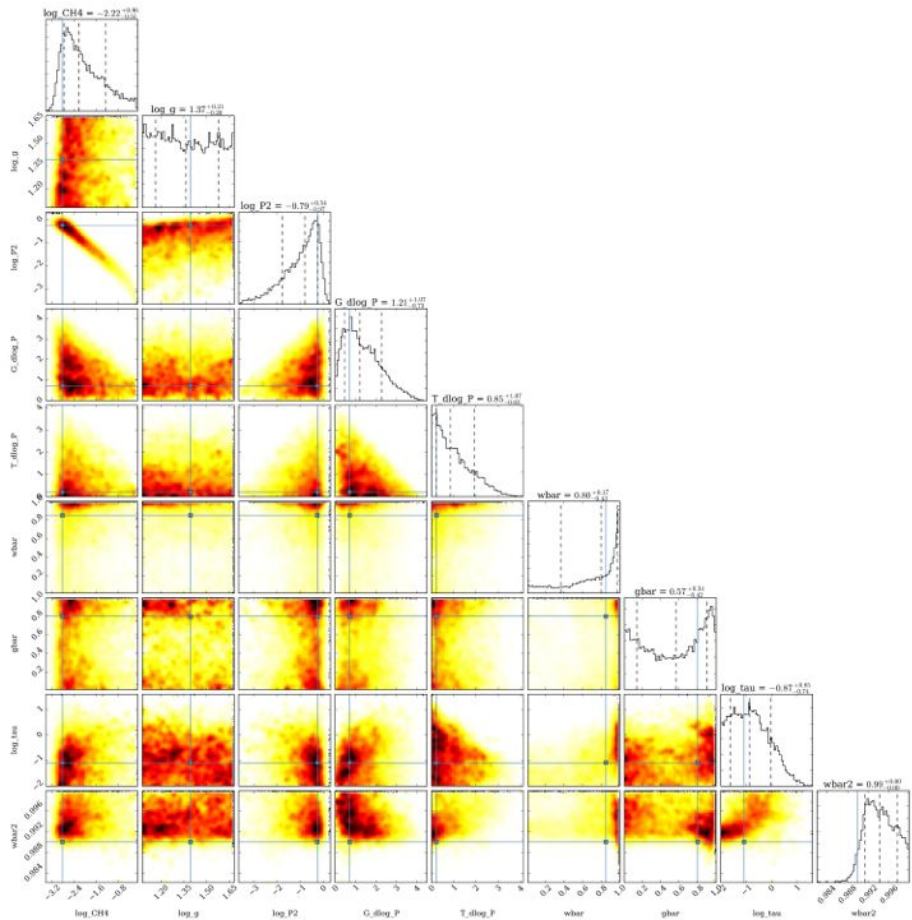

**Figure 24** Posterior probability distributions for the albedo model parameters, as sampled by the MCMC algorithm, for the Jupiter albedo with simulated data for SNR=10, noise correlation length 25 nm, and gravity constrained to lie within a factor of 2 of the true value. The distributions contain 324000 samples, after excluding the burn-in phase for the MCMC chain. The darker regions represent higher probability. The marginalized probability distributions for each parameter are shown on the diagonal. The error bars and the expectation value are calculated to indicate the 16%,

84% and 50% quantiles, respectively, such that the entire confidence interval contains 68% of the values. An expanded view of the methane retrieval is on the cover page.

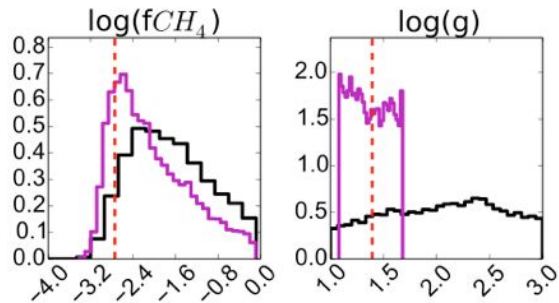

**Figure 25.** Marginal posterior probability distributions for the methane abundance (left) and the surface gravity (right) of Jupiter, as derived from the simulated *WFIRST-AFTA* data. The purple line corresponds to the case where we set a prior limiting the surface gravity to a small interval around the real value (+/- a factor of 2). In this case, the posterior probability for f$CH_4$ becomes strongly peaked at the actual measured value for Jupiter (vertical red dashed line). This shows that the width of this distribution is mainly due to the correlation between f$CH_4$ and *g*.

**Future Improvements**
Because of the nature of the quick study we had to make a number of simplifications to the analysis to make our task practicable. However future work should address these simplifications and address their roles in the fidelity of the retrievals. Foremost among those that should be explored include: planetary radius uncertainty, thermal profile uncertainty, and orbital phase uncertainty. In addition the retrieval of more atmospheric abundances should be explored, particularly water and alkali gasses. In addition a somewhat more general cloud model should be adopted.

# CONCLUSIONS

We have used an MCMC retrieval method to quantify the confidence intervals on the atmospheric methane abundance and cloud structure, using a simple atmospheric model with the methane abundance, surface gravity, and 7 cloud properties as free parameters. Given the limited time available we were not able to optimize the modeling framework. Nevertheless we find that reflected light spectra of the quality expected from a space based direct imaging exoplanet mission is sufficient to place interesting constraints on important planetary atmosphere characteristics, particularly methane mixing ratio and, in some cases, cloud albedo. The MCMC method is powerful for determining correlations among parameters and identifying which ones are unconstrained by the data. In this case demonstrating the value in the synthetic datasets, even at low signal to noise ratios.

# Appendix I: Noise Model

Our final noise model follows the prescription on the SFOM_4Dec2013c document, as provided by Wesley Traub. All relevant equations have been reproduced below from this document.

We define the signal-to-noise (SNR0) as corresponding to the integrated number of counts in a 10%-wide bandpass centered at 550 nm. This value was set to 5, 10, and 20 for our simulation runs.

$$\text{signal (elec)} = n_{pl} \times t \quad \text{(within the FWHM of the planet image)}$$

$$\text{noise (elec)} = [\, n_{total} \times t + (f_{pp} \times n_{rawspeckle} \times t)^2 \,]^{1/2}$$

$$n_{total}(\text{elec/s}) = [\, n_{pl} + n_{zodi} + n_{rawspeckle} + D_c \times m_{pix} + CIC \times m_{pix}/t_{frame} \,] \times ENF^2 + (N_R/G)^2 \times m_{pix}/t_{frame}$$

$$\text{signal/noise} = SNR_0$$

$$\boxed{\; t\,(\sec) = \frac{SNR_0^2 \times n_{total}\,(e/s)}{n_{pl}^2 - (SNR_0 \times f_{pp} \times n_{rawspeckle})^2} \;}$$

$$n_{total} = [\, n_{pl} + n_{zodi} + n_{spec} \times (1 + f_{pp}) + D_c \times m_{pix} + CIC \times m_{pix}/t_{frame} \,] \times ENF^2$$
$$+ (N_R/G)^2 \times m_{pix}/t_{frame} \quad (\text{elec/s})$$

Our wavelength grid covers the 0.6-1.0 μm bandpass with a resolution R=70.

Starting from the planetary albedo and the stellar spectrum as a blackbody at 6000 K, we have calculated the expected number of photons in each spectral bin. This number was converted to a count rate, using the "typical values" given in the table below. For all cases we have considered the detector to be EMCCD.

| parameter | units | CCD typical value | EMCCD typical value |
|---|---|---|---|
| $n_{pl}$ | elec/sec | 0.012 | 0.012 |
| $n_{zodi}$ | elec/sec | 0.012 | 0.012 |
| $n_{spec}$ | elec/sec | 0.010 | 0.010 |
| $m_{pix}$ | pixels | 5 | 5 |
| $D_c$ | elec/(pixel sec) | 0.001 | 0.001 |
| $N_R$ | RMS elec/(pixel frame) | 3 | 3 |
| $t_{frame}$ | sec | 300 | 300 |
| CIC | elec/(pixel frame) | 0 | 0.001 |
| ENF | | 1 | 1.414 |
| G | | 1 | 1000 |
| t | sec | 33,000 | 14,000 |

We calculate the integration time necessary to obtain a SNR0 of 5, 10, or 20, as defined above. The total number of counts on the detector is given by the planet counts, the speckle noise counts, the zodiacal light (zodi) and the detector background summing all other sources. The integration time is then used to scale the signal and noise across the bandpass in every channel, by calculating the expected number of counts.

The observed spectrum is simulated assuming that the planet and zodi counts have a Poisson distribution (per channel), while the speckle and detector noise counts have a Gaussian distribution. Since the Poisson-distributed counts are uncorrelated, the correlated noise will only affect the Gaussian-distributed counts. The total noise contribution of the Gaussian-distributed counts is split into 2 components, one correlated, and one uncorrelated. The noise correlations are also considered gaussian, with a length scale of either 25 or 100 nm. This noise is generated as a Gaussian random process with a squared-exponential kernel (basically, a covariance matrix). The contributions of both correlated and uncorrelated components are set to be equal (i.e. they have equal variances). The final error bars are computed individually for each simulated data point.

These parameter combinations result in 6 simulated datasets for each planet model:
- SNR=5 with correlations = 25 nm or 100 nm,
- SNR=10 with correlations = 25 nm or 100 nm,
- SNR=20 with correlations= 25 nm or 100 nm.
Although our simulated data cover a wider range, the model fits only cover the 0.6-1.0 µm range specified in the instrument description.

The following plots exemplify the simulated data for a Jupiter-like planet around a Sun-like star, at a distance of 25 pc from our Solar System.

SNR=5, corr_len=25

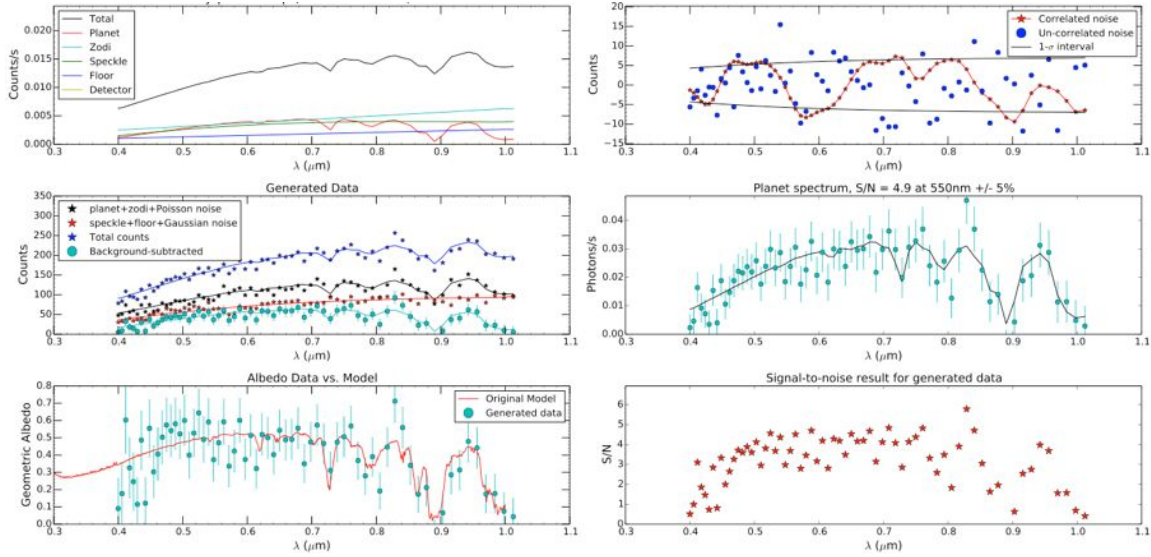

SNR=5, corr_len=100

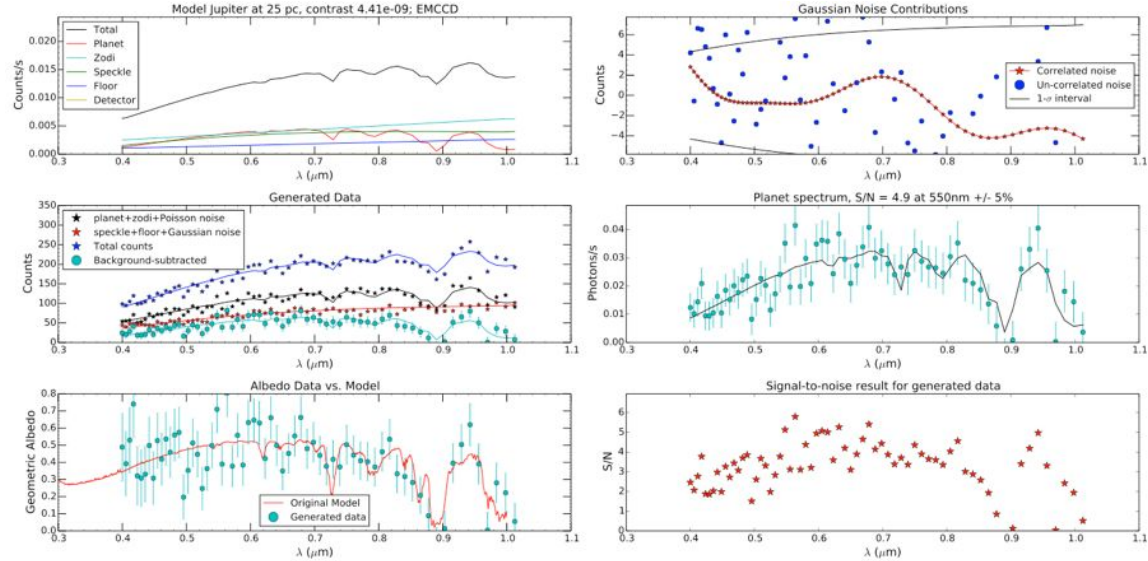

**Figure A1.** In each set, the panels are as follows:
*Top left* – expected count rates from all different sources: planet–red, zodi-cyan, speckle-green, detector noise–blue and yellow (too small to see). Total count rate-black.
*Middle left* – total number of counts after calculating the integration time needed to get a SNR of 5. The model counts are solid lines, and the simulated data are stars.
*Top right* – correlated and un-correlated noise contributions. These are added-in when generating the red star data points in the middle-left panel.
*Middle right* – simulated data converted to photon rate, after background subtraction (cyan), compared to the input model (black).
*Bottom left* – simulated data converted back to geometric albedo, after division by the stellar spectrum (cyan), vs actual albedo of Jupiter (red).
*Bottom right* – SNR of the simulated data in each wavelength bin. The nominal SNR (5) corresponds to a 10% band around 550 nm. Note individual variation from ~0 to ~6.

SNR=10, corr_len=25

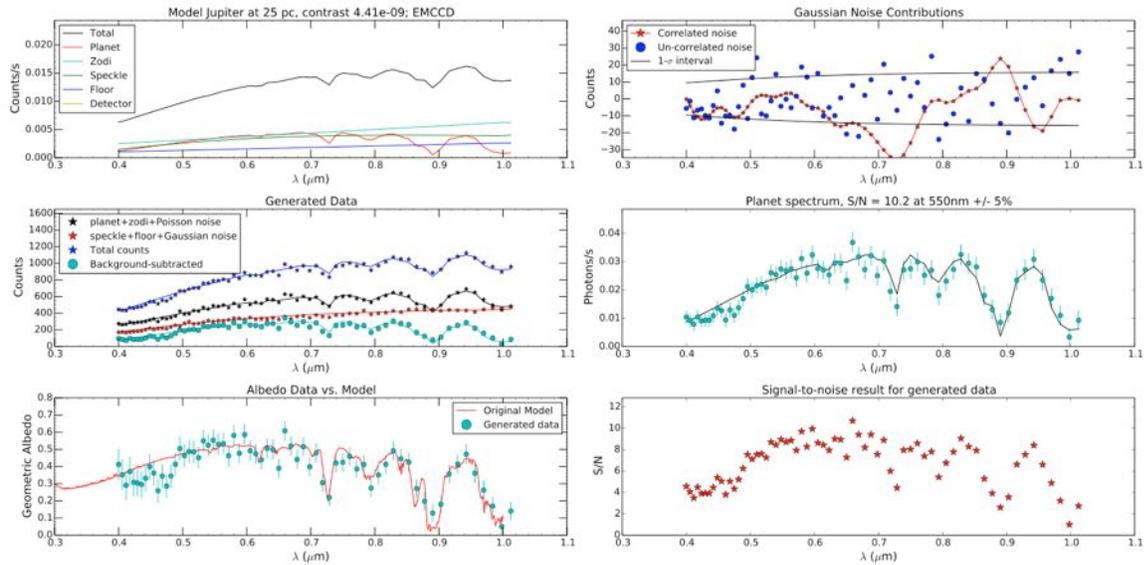

SNR=10, corr_len=100

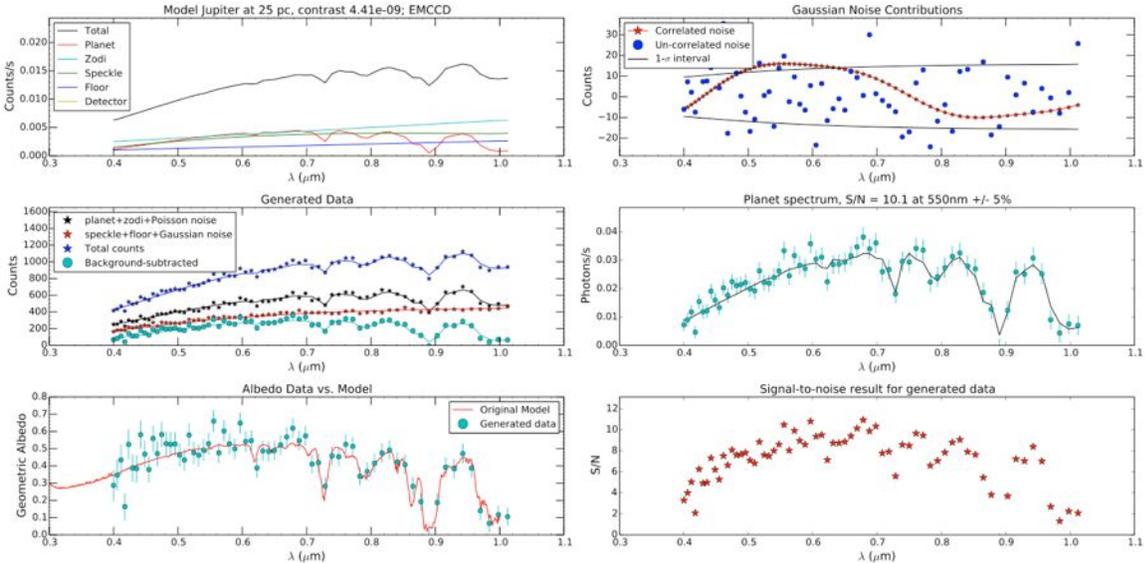

**Figure A2.** In each set, the panels are as follows:
*Top left* – expected count rates from all different sources: planet–red, zodi-cyan, speckle-green, detector noise–blue and yellow (too small to see). Total count rate is shown in black.
*Middle left* – total number of counts after calculating the integration time needed to get a SNR of 10. The model counts are solid lines, and the simulated data are stars.
*Top right* – correlated and un-correlated noise contributions. These are added-in when generating the red star data points in the middle-left panel.
*Middle right* – simulated data converted to photon rate, after background subtraction (cyan), compared to the input model (black).
*Bottom left* – simulated data converted back to geometric albedo, after division by the stellar spectrum (cyan), vs actual albedo of Jupiter (red).
*Bottom right* – SNR of the simulated data in each wavelength bin. The nominal SNR

(10) corresponds to a 10% band around 550 nm. Note individual variation from ~1 to ~11.

SNR=20, corr_len=25

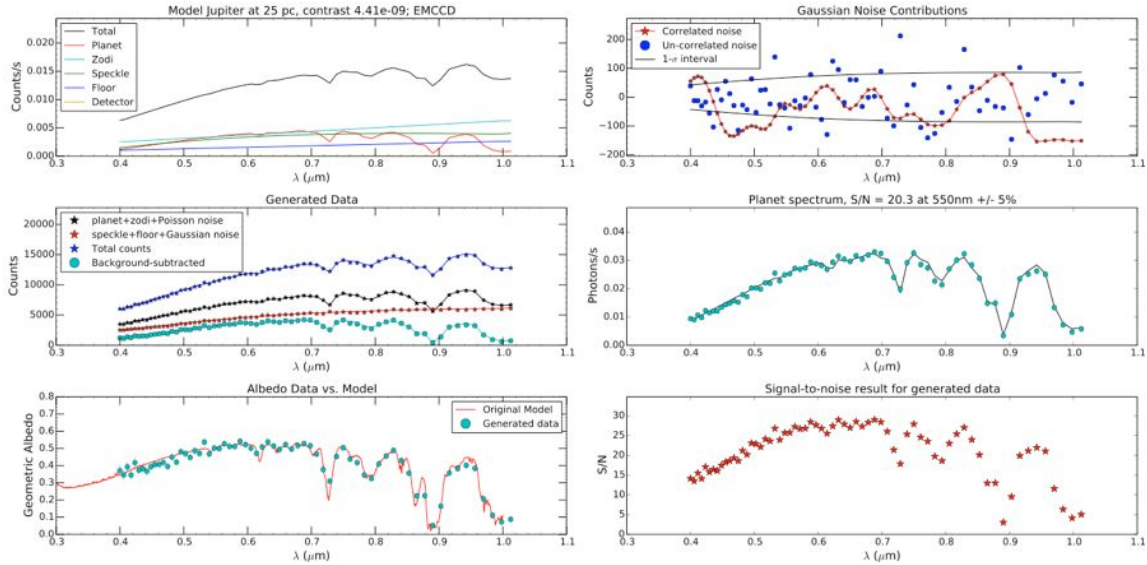

SNR=20, corr_len=100

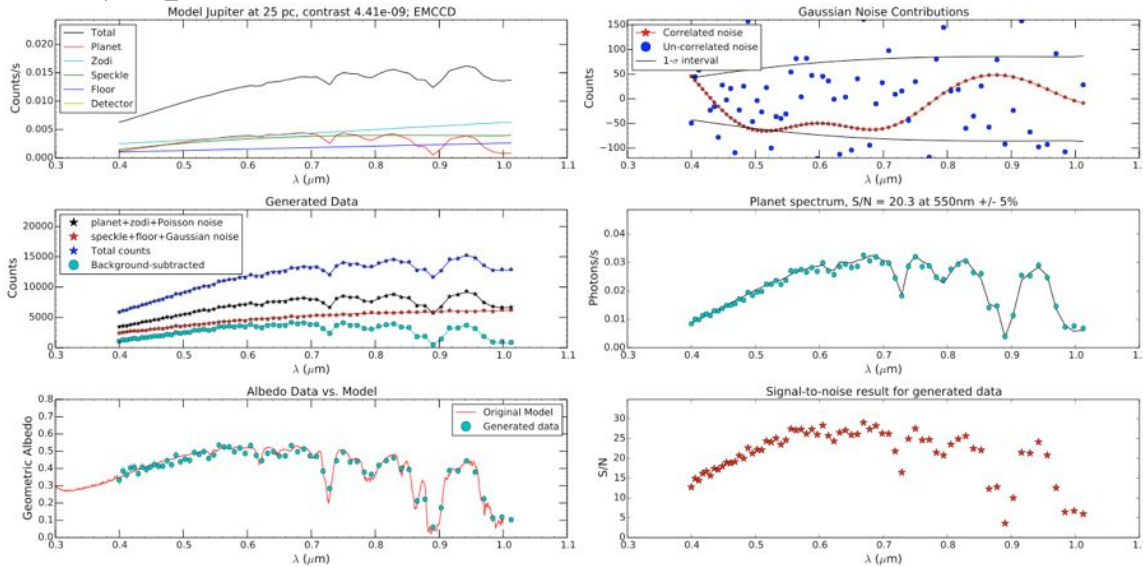

**Figure A3.** In each set, the panels are as follows:

*Top left* – expected count rates from all different sources: planet –red, zodi- cyan, speckle-green, detector noise – blue and yellow (to small to see). The total count rate is shown in black.

*Middle left* – total number of counts after calculating the integration time needed to get a SNR of 20. The model counts are solid lines, and the simulated data are stars.

*Top right* – correlated and un-correlated noise contributions. These are added-in when generating the red star data points in the middle-left panel.

*Middle right* – simulated data converted to photon rate, after background subtraction (cyan), compared to the input model (black).

*Bottom left* – simulated data converted back to geometric albedo, after division by the stellar spectrum (cyan), vs actual albedo of Jupiter (red).
*Bottom right* – SNR of the simulated data in each wavelength bin. The nominal SNR (20) corresponds to a 10% band around 550 nm. Note individual variation from ~3 to ~30.

# ACKNOWLEDGEMENT

A study outlining the scientific case for a high-contrast coronagraph on WFIRST in support of its exoplanet campaign, performed on behalf of the WFIRST/AFTA Science Definition Team and the Exo-S and Exo-C Science and Technology Definition Teams.